\documentclass[letterpaper,twocolumn,prl,aps,superscriptaddress,amsmath,amssymb,floatfix]{revtex4-1}
\usepackage{mathptmx}
\usepackage[latin9]{inputenc}
\usepackage{color}
\usepackage{verbatim}
\usepackage{float}
\usepackage{amsmath}
\usepackage{amssymb}
\usepackage{graphicx}
\usepackage{esint}

\makeatletter

\usepackage{textcomp}
\usepackage{epstopdf}

\pdfpageheight\paperheight
\pdfpagewidth\paperwidth

\providecommand{\tabularnewline}{\\}

\@ifundefined{textcolor}{}{%
 \definecolor{BLACK}{gray}{0}
 \definecolor{WHITE}{gray}{1}
 \definecolor{RED}{rgb}{1,0,0}
 \definecolor{GREEN}{rgb}{0,1,0}
 \definecolor{BLUE}{rgb}{0,0,1}
 \definecolor{CYAN}{cmyk}{1,0,0,0}
 \definecolor{MAGENTA}{cmyk}{0,1,0,0}
 \definecolor{YELLOW}{cmyk}{0,0,1,0}
}

\usepackage{xcolor}
\usepackage{soul}
\definecolor{blue}{rgb}{0,0,1}
\definecolor{red}{rgb}{1,0,0}
\definecolor{green}{rgb}{0,1,0}

\usepackage{soul}

\usepackage[unicode=true,
 bookmarks=true,bookmarksnumbered=false,bookmarksopen=false,
 breaklinks=false,pdfborder={0 0 1},backref=false,colorlinks=true]
 {hyperref}
\hypersetup{
 linkcolor=magenta,urlcolor=blue,citecolor=blue,pdfstartview={FitH},hyperfootnotes=false}

\makeatother

\begin{document}

\address{Laboratory of Quantum Information, University of Science
and Technology of China, Hefei, Anhui 230026, China}
\address{Center for Quantum Information, Institute for Interdisciplinary Information
Sciences, Tsinghua University, Beijing 100084, China}
\address{School of Civil Engineering, Hefei University of Technology,
Hefei 230009, China}
\address{Anhui Province Key Laboratory of Quantum Network, University of Science and Technology of China, Hefei 230026, China}
\address{CAS Center For Excellence in Quantum Information and Quantum
Physics, University of Science and Technology of China, Hefei, Anhui
230026, China}
\address{Hefei National Laboratory, Hefei 230088, China}


\title{Proposal of cavity quantum acoustodynamics platform based on Lithium Niobate-on-Sapphire chip}

\author{Xin-Biao Xu}
\address{Laboratory of Quantum Information, University of Science
and Technology of China, Hefei, Anhui 230026, China}
\address{Anhui Province Key Laboratory of Quantum Network, University of Science and Technology of China, Hefei 230026, China}

\author{Lintao~Xiao}
\address{Center for Quantum Information, Institute for Interdisciplinary Information
Sciences, Tsinghua University, Beijing 100084, China}

\author{Bo~Zhang}
\address{Center for Quantum Information, Institute for Interdisciplinary Information
Sciences, Tsinghua University, Beijing 100084, China}

\author{Weiting~Wang}
\email{wangwt2020@tsinghua.edu.cn}
\address{Center for Quantum Information, Institute for Interdisciplinary
Information Sciences, Tsinghua University, Beijing 100084, China}

\author{Jia-Qi Wang}
\address{Laboratory of Quantum Information, University of Science
and Technology of China, Hefei, Anhui 230026, China}
\address{Anhui Province Key Laboratory of Quantum Network, University of Science and Technology of China, Hefei 230026, China}

\author{Yu Zeng}
\address{Laboratory of Quantum Information, University of Science
and Technology of China, Hefei, Anhui 230026, China}
\address{Anhui Province Key Laboratory of Quantum Network, University of Science and Technology of China, Hefei 230026, China}

\author{Yuan-Hao Yang}
\address{Laboratory of Quantum Information, University of Science
and Technology of China, Hefei, Anhui 230026, China}
\address{Anhui Province Key Laboratory of Quantum Network, University of Science and Technology of China, Hefei 230026, China}

\author{Bao-Zhen Wang}
\address{School of Civil Engineering, Hefei University of Technology,
Hefei 230009, China}

\author{Xiaoxuan~Pan}
\address{Center for Quantum Information, Institute for Interdisciplinary Information
Sciences, Tsinghua University, Beijing 100084, China}

\author{Guang-Can Guo}
\address{Laboratory of Quantum Information, University of Science
and Technology of China, Hefei, Anhui 230026, China}
\address{Anhui Province Key Laboratory of Quantum Network, University of Science and Technology of China, Hefei 230026, China}
\address{CAS Center For Excellence in Quantum Information and Quantum
Physics, University of Science and Technology of China, Hefei, Anhui
230026, China}
\address{Hefei National Laboratory, Hefei 230088, China}

\author{Luyan Sun}
\address{Center for Quantum Information, Institute for Interdisciplinary
Information Sciences, Tsinghua University, Beijing 100084, China}
\address{Hefei National Laboratory, Hefei 230088, China}

\author{Chang-Ling Zou}
\email{clzou321@ustc.edu.cn}
\address{Laboratory of Quantum Information, University of Science
and Technology of China, Hefei, Anhui 230026, China}
\address{Anhui Province Key Laboratory of Quantum Network, University of Science and Technology of China, Hefei 230026, China}
\address{CAS Center For Excellence in Quantum Information and Quantum
Physics, University of Science and Technology of China, Hefei, Anhui
230026, China}
\address{Hefei National Laboratory, Hefei 230088, China}

\date{\today}

\begin{abstract}
A scalable hybrid cavity quantum acoustodynamics (QAD) platform is proposed. The architecture integrates superconducting transmon qubits with phononic integrated circuits on a single chip made by lithium niobate-on-sapphire substrate. The platform supports tightly confined and guided phononic modes in unsuspended waveguides and microring structures, while the superconducting qubits reside on the sapphire substrate. Efficient piezoelectric coupling between the phononic modes and transmon qubits can be achieved using interdigital transducers as part of the qubit's shunt capacitance. Numerical calculations demonstrate the feasibility of achieving strong coupling between the phononic microring resonator and the transmon qubit. This hybrid cavity QAD platform opens up new opportunities for quantum information processing and the study of novel quantum acoustic phenomena.
\end{abstract}

\maketitle

\section{Introduction}

The coherent coupling between quantized acoustic wave (phonon) excitations and artificial atoms in solids have emerged as a fascinating research field~\cite{Schuetz2015,Manenti2017,Kuruma2025,Odeh2025}, offering new opportunities for studying fundamental physics and advancing quantum technologies. As an analogue to the cavity and waveguide quantum electrodynamics for studying photon-atom interaction in quantum optics, the phonon-qubit interaction contributes to a potential research direction of cavity and waveguide quantum acoustodynamics (QAD)~\cite{Zou2025}. The cavity QAD effects have been studied in various qubit systems, including quantum dots~\cite{DeCrescent2022_aqd}, ions~\cite{DeCrescent2022_aion}, NV centers~\cite{Golter2016a_anv}, silicon vacancy center~\cite{Kuzyk2018_asiv,Maity2020_asiv} and superconducting qubits~\cite{Gustafsson2014_surf}, providing alternative approaches to the control and manipulation of quantum states through the phonons. Remarkable achievements have been demonstrated, including single-phonon control, phonon-mediated quantum gates, and macroscopic quantum states with phononic modes. To date, most QAD studies have focused on platforms such as surface acoustic wave devices~\cite{Gustafsson2014_surf,Bienfait2019_surface,Satzinger2018_surface}, bulk acoustic wave devices~\cite{Chu2017_bulk,VonLupke2024_CQAD}, or suspended mechanical devicess~\cite{Wollack2022_suspended}. While these systems have yielded significant insights, they face challenges in terms of scalability, fabrication complexity, and the ability to confine and route acoustic waves efficiently.

In recent years, the development of phononic integrated circuits (PnICs) has opened up new possibilities for further exploring cavity and waveguide QAD. PnICs, which are analogous to integrated photonic circuits~\cite{Zhu2021_LNO}, enable the efficient manipulation and routing of acoustic waves in complex, chip-scale devices~\cite{Fu2019,Wang2020,Mayor2021,Shao2022,Shao2022a}. By leveraging high-acoustic-velocity-contrast materials, PnICs can confine acoustic fields to wavelength-scale dimensions, greatly enhancing the interaction between phonons and matter~\cite{Xu2022,Mayor2021,Bicer2022,Feng2023}. This novel platform offers a scalable and versatile approach to designing and realizing functional phononic devices, such as acoustic beam splitters~\cite{Feng2023}, Mach-Zehnder interferometers~\cite{Shao2022,Shao2022a}, mode converters~\cite{Wang2022}, and microring resonators~\cite{Xu2022}. With the ability to assemble these building blocks into larger, more complex circuits, PnICs provide an exciting avenue for investigating quantum acoustodynamics and developing practical applications in quantum information processing and sensing~\cite{Hann2019}.

\begin{figure}[h]
\centering{}
\includegraphics[width=1\columnwidth]{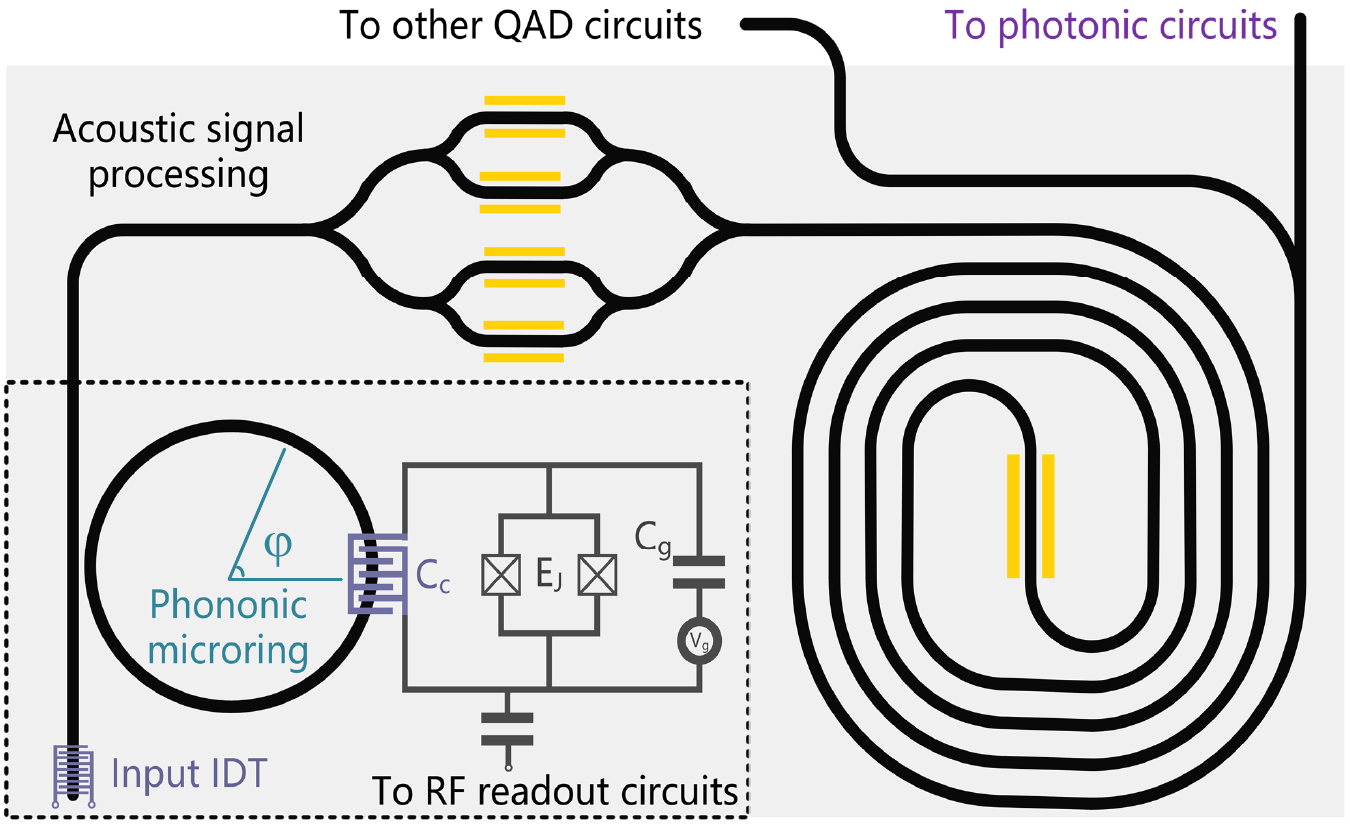}
\caption{\label{fig:1}
Hybrid cavity quantum acoustodynamics (QAD) platform architecture on lithium niobate-on-sapphire (LNoS). The platform integrates phononic integrated circuits (PnIC) with superconducting quantum circuit on a single chip. The PnIC consists of phononic waveguides and microring resonators, which support tightly confined propagating phononic modes at GHz frequencies. Dahsed box: superconducting transmon qubits and relevant superconducting components can couple with the phononic mode through interdigital transducer (IDT), which is designed as part of the transmon's shunt capacitance. The azimuthal angle $\varphi$ indicates the position along the microring resonator. }
\end{figure}

In this work, we propose a hybrid cavity QAD platform that integrates PnICs on a lithium niobate-on-sapphire (LNoS) chip with superconducting artificial atoms, namely transmon qubits~\cite{Clarke2008_qubit,Krantz2019_qubit}. We first analyze the basic components of the platform, including phononic waveguides, microring resonators, and interdigital transducers (IDTs) for coupling between qubits and acoustic waves. As strong coupling is a prerequisite for further development of the platform, we theoretically investigate the interaction strength between GHz-frequency acoustic modes in the microring resonator and the transmon qubit, demonstrating that strong coupling is achievable in this hybrid system. Additionally, we examine the impact of the anisotropic properties of lithium niobate and sapphire on the coupling strength. Lastly, considering the potential fabrication imperfections in realistic nanoscale devices, we analyze the effect of variations in the IDT duty cycle on the coupling strength, particularly for high-frequency IDTs operating at around 6 GHz. Our work establishes the feasibility and potential of the LNoS-based hybrid cavity QAD platform, paving the way for experimental realizations and further explorations of QAD in integrated phononic systems.

\section{Hybrid cavity QAD architecture}

Figure~\ref{fig:1} illustrate the schematics of the proposed cavity QAD architecture, which is scalable by integrating both PnICs and superconducting circuits. To realize the coherent interface between the two circuits without disrupting the performance of both, the materials of the architecture need to meet the following requirements:

(i) The waveguide layer of PnICs needs be a good piezoelectric material to enable the efficient conversion between microwave and acoustic wave.

(ii) The speed of acoustic wave in waveguide layer should be slower then that in substrate for strong confinement of phononic modes in the devices.

(iii) Both the waveguide layer and the substrate should be singe-crystal and exhibit low loss at both microwave and acoustic domains.

To further illustrate the variety of candidate materials for realizing the hybrid QAD architecture, Table~\ref{tab:material_properties_updated} surveys a list of potential materials for waveguide layer and substrate~\cite{Gualtieri1994_piez}. The waveguide layer materials, such as lithium niobate (LiNbO$_3$, LN), lithium tantalate (LiTaO$_3$), and zinc oxide (ZnO), are chosen for their low acoustic velocities, which enable tight confinement of phonons and strong phonon-qubit interactions. On the other hand, the substrate materials, including sapphire (Al$_2$O$_3$), silicon (Si), and diamond, exhibit high acoustic velocities, ensuring tight phonon confinement. The table provides both transverse and longitudinal acoustic wave velocities for each material, as both types of waves can be employed in hybrid QAD circuits depending on the specific design requirements. This wide selection of materials enables researchers to tailor their hybrid QAD systems to specific applications and performance requirements, paving the way for the development of advanced quantum devices and circuits. Among various potential combinations of materials, the LNoS proivdes an excellent candidate for the hybrid circuits. LN combines exceptional piezoelectric coupling coefficients with low acoustic velocity (about 3500 $\mathrm{m/s}$), while sapphire provides a high-velocity substrate (about 5800 $\mathrm{m/s}$) with ultra-low dielectric loss as it has been widely used in superconducting integrated circuits. Additionally, the LN holds excellent electro-optical and second-order nonlinear coefficients and has been extensively studied for photonics applications~\cite{Zhu2021_LNO}.

\begin{table}[t]
\centering
\caption{Properties of potential materials for hybrid QAD circuits. Both transverse ($V_{t}$) and longitudinal ($V_{l}$) acoustic wave velocities are provided.}
\label{tab:material_properties_updated}
\begin{tabular}{|c|c|c|c|c|}
\hline
Material & $V_{l}$ (m/s) & $V_{t}$ (m/s) & Piezoelectricity  \tabularnewline
\hline
\hline
$\mathrm{LiNbO\ensuremath{_{3}}}$~\cite{Warner1967_LN} & 6572 & 3573 & Yes\tabularnewline
\hline
$\mathrm{LiTaO\ensuremath{_{3}}}$~\cite{Warner1967_LN} & 5592  & 3552 & Yes\tabularnewline
\hline
$\mathrm{ZnO}$~\cite{Azuhata2003_zno,Sun2020} & 5790 & 2700 & Yes\tabularnewline
\hline
$\mathrm{GaN}$~\cite{levinshtein2001properties_GaN} & 7350 & 4578 & Yes\tabularnewline
\hline
$\mathrm{AlN}$~\cite{levinshtein2001properties_GaN} & 10169 & 6369 & Yes\tabularnewline
\hline
$\mathrm{Al_{1-x}Sc_{x}N}$(x=0.12)~\cite{Park2020_alscn} & 10144 & 5559 & Yes\tabularnewline
\hline
$\mathrm{Quartz}$~\cite{Pohl2002_quartz} & 5700 & 3158 & Yes\tabularnewline
\hline
$\mathrm{4H\,SiC}$~\cite{Zhang2020_sic} & 12493 & 7126 & Yes\tabularnewline
\hline
$\mathrm{Sapphire}$~\cite{auld1973acoustic_sapphire} & 10658 & 5796 & No\tabularnewline
\hline
$\mathrm{Si}$~\cite{Hopcroft2010_silicon} & 8433 & 5843 & No\tabularnewline
\hline
$\mathrm{Diamond}$~\cite{Gualtieri1994_diamond} & 18000 & 12000 & No\tabularnewline
\hline
\end{tabular}
\end{table}

The quantum interface between the phononic devices and superconducting qubit is realized through the IDT, which contributed to the transmon's shunt capacitance, as shown in the dashed box in Fig.~\ref{fig:1}. The piezoelectric coupling between the traveling wave phononic microring resonator and the transmon can be enhanced by increasing the length of the IDT region, with the period of the coupling IDT matching that of the acoustic modes. The coupled phonon-qubit system can be characterized by either readout the superconducting circuits~\cite{Devoret2013_Superconductingcircuits} or detect the acoustic transmittance of the phononic microring through IDT~\cite{Dahmani2020_IDT}. Alternatively, the acoustic signal in the microring can also be directed to PnIC for further information processing and or guided to other superconducting devices through phononic waveguide. Both the localized (microring resonator) and itinerant (waveguide) phonons contributes the basic building blocks for scalable quantum devices for further single-phonon manipulation and functional long-range qubit entanglement experiments.

\section{Phononic integrated circuits}

\subsection{The Phononic Waveguide}

\begin{figure}[t]
\centering{}\includegraphics[width=1\columnwidth]{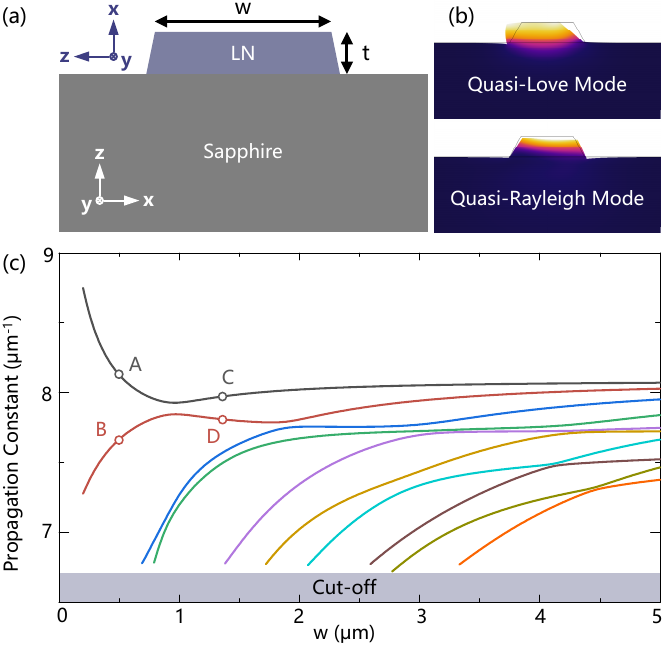}\caption{\label{fig:2}
Phononic waveguide design and modal characteristics. (a) Cross-section view of straight waveguide, where a X-cut LN layer (width $w$, thickness $t$) is on a c-plane sapphire substrate. (b) Normalized displacement fields of fundamental quasi-Love and quasi-Rayleigh modes at 6\,GHz, with $w=500\,\mathrm{nm}$ and $t=200\,\mathrm{nm}$. of the bus waveguide. (c) Dispersion diagram showing propagation constants of phononic modes as functions of the waveguide width $w$. Points A and B indicate $w=500\,\mathrm{nm}$, and points C and D indicate $w=1400\,\mathrm{nm}$.}
\end{figure}

Phononic waveguide is the most basic building block of PnICs. Figure~\ref{fig:2}(a) shows the cross-section of the bus waveguide coupled to a microring resonator in our LNoS platform. We employ c-plane saphhire as the substrate, and $X$-cut LN for the waveguide layer. The acoustic waveguide is designed to propagated along the $Y$ axis for maximazing the piezoelectric strain coefficient of LN ($d_{15}\approx74\,\mathrm{pC/N}$). Consequently, the quasi-Love mode can be excited with high transduction efficiency by the input IDT, as shown in Fig.~\ref{fig:1}. Figure~\ref{fig:2}(b) shows the fundamental quasi-Love and quasi-Rayleigh mode of the waveguide with $w=500$\,nm and $t=200$\,nm. It can be found that the both phononic waveguide modes are tightly confined in the waveguide, with negligible evanscent acoustic field in the substrate. The modes are also characterized by the dispersion curves, as shown in Fig.~\ref{fig:2}(c), showing the dependence of propagation constant on the waveguide width. We found multiple avoided-crossing features in the dispersion curves, indicating the coupling between different phononic modes when their propagation constants coincident, with the coupling mainly originate from the material anisotropic and tight confinement. For example, we identify that points A and D are quasi-Love modes, while B and C are quasi-Rayleigh modes. In the intermediate width range (between A and C), the eigenmodes become hybridization of quasi-Love and quasi-Rayleigh modes~\cite{Wang2022}.

Several competing factors should be considered when design the waveguides for PnIC. To optimize the IDT transduction efficiency, the waveguide should operate in the geometry parameter regimes where the mode hybridization is suppressed. A narrow waveguide reduces the propagation constant for most modes, thus increase the portion of energy in substrate. Thus, the weaker confinement with narrow waveguide allows enhanced evanescent field coupling between adjacent phononic waveugide for realizing multiport devices, while suffering from increase scattering losses due to the sidewall roughness. These trade-offs suggests the careful selection of waveguide geometry parameters for specific devices.

\subsection{The Phononic Microring Resonator}

\begin{figure}[b]
\centering{}\includegraphics[width=1\columnwidth]{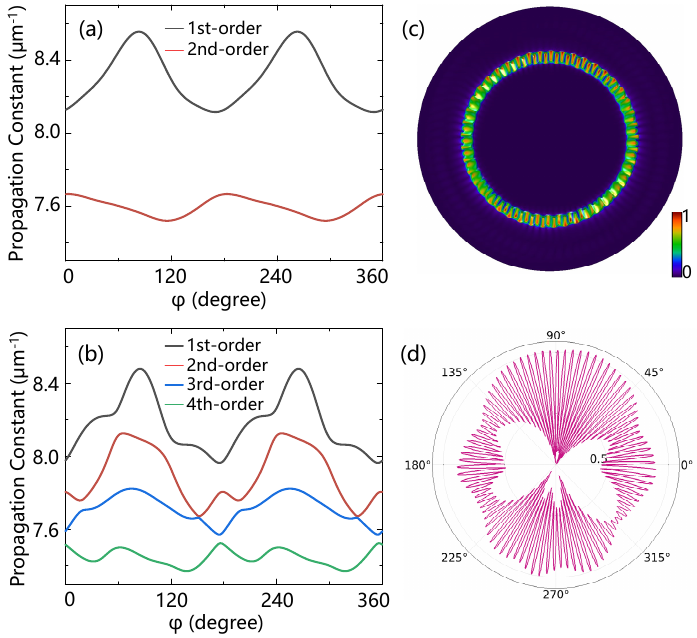}
\caption{\label{fig:3} Phononic modes in microring resonators. (a)-(b) The propagation constant of the eigenmodes in waveguide versus different $\varphi$ for narrow waveguide (width $w=500\,\mathrm{nm}$, Point A in Fig.~2) and wide waveguide (width $w=1400\,\mathrm{nm}$, point C in Fig.~2), respectively. (c) The displacement field of eigenmode in a microring resonator with a radius of $5\,\mathrm{\mu m}$. (d) The azimuthal angular ($\varphi$) dependence of in-plane displacement component along the microring. }
\end{figure}

Phononic microring resonators can be designed by simply bending the waveguide, but challenge arised due to crystal anisotropy. While Fig.~\ref{fig:2}(c) indices that mode hybridization at points A and D is negligible for straight waveguide alaong the Y-axis, the situation for microring becomes drastically different. It is because that for a loop of waveguide, the crystal orientation of both LN and sapphire are changed at different azimuth angle $\varphi$. For waveguide with fixed width and etch depth, the mode hybridization may be found at specific azimuth angles. Figure~\ref{fig:3}(a) shows the propagation constant of the phononic modes in waveguide at different $\varphi$ with the waveguide width at A (B) point. For narrow waveguide width, there are only two eigenmodes be well-confined in the waveguide, while  the other modes are cut-off. The curves exhibit a period of approximately 180 degrees, which agree with the period of 180 degrees for the elastic coefficients of X-cut LN. Although the substrate have a different symmetry, the anisotropic of LN dominates because the energy of phononic eigenmodes are well confined in the LN layer. Besides, there are no avoid-crossing regions in the curves, indicating that the quasi-Love and quasi-Rayleigh modes maintain their polarization purity when propagating along the microring.

However, the situation changes significantly for wider waveguides. As we can see from Fig.~\ref{fig:3}(b), when waveguide width is at point C(D), there are four eigenmodes and many avoid-crossing regions exist. For each curves, the mode field will change at different $\varphi$. So the displacement field at different $\varphi$ of the eigenmode of the microring may be different. To visualize this effect, Fig.~\ref{fig:3}(c) presents the numerical results of the displacement field for one of the eigenmodes in a microring. Due to the limitation of computational resource, we only calculate the micoring of a radius of $5\,\mu\text{m}$ as an exmpale. Figure~\ref{fig:3}(d) provides the corresponding magnitude of the in-plane displacement as a function of the azimuthal angle $\varphi$ at the centerline position of the microring waveguide. The numerical results reveal non-uniform mode profile in the microring: the displacement field is predominantly in-plane (quasi-Love mode) at the top and bottom of the microring ($\varphi=0^\circ,\,180^\circ$), while transitions to out-of-plane motion dominate (quasi-Rayleigh) at the sides ($\varphi=90^\circ,\,270^\circ$). Consequently, the coupling strength between the transmon qubits and phononic microring in Fig.~\ref{fig:1} will show significant dependence on the $\varphi$, as the piezoelectric interaction tensor is related to the crystal orientation and the mode profile is not uniform along the azimuthal direction. Therefore, we select the narrow waveguide width (point A) to maintain polarization purity throughout the ring, thus simply the structure design and calculations in the following.

\subsection{The Coupling IDT }

\begin{figure}[t]
\centering{}\includegraphics[width=1\columnwidth]{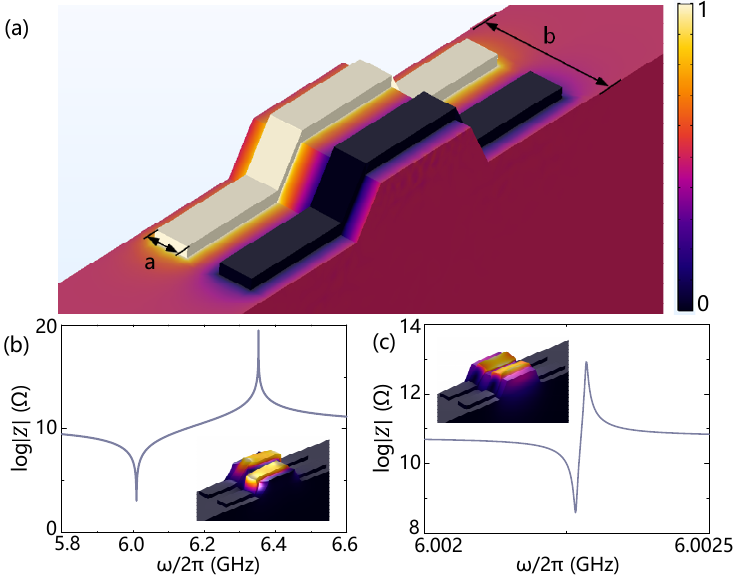}\caption{\label{fig:4} (a) The electric potential distribution of single finger pair of the coupling IDT with  $1V$ voltage is applied between the metal strips. $a$ is the width of the metal strips, $b$ is the width of a finger pair. The duty ratio of IDT is $\eta=a/b$. (b) and (c) are the electrical impedance of the IDT versus frequency for quasi-Love and quasi-Rayleigh mode excitation with waveguide along $Y$ axis. The insets show the deformation of the IDT at the resonant frequency, which is adjusted to $6\,\mathrm{GHz}$ with $\eta=0.5$.}
\end{figure}

The IDT provides the critical quantum interface in our architecture for high-efficiency bidirectional conversion between propagating phonons in the waveguide and the microwave photons in superconducting circuit~\cite{Mamishev2004_IDT}. Figure~\ref{fig:4}(a) illustrates our IDT design, showing the electric potential distribution of single finger pair of the coupling IDT when $1\,\mathrm{V}$ bias voltage is applied. The coupling strength is maximized when the electric field distribution generated by the IDT matches that generated by the acoustic wave, similar to the phase matching condition in optics. Considering the width of the LN waveguide is only 500\,nm, we set the metal strips of IDT across the LN waveguide directly to reduce the fabrication complexity in practice. The IDT geometry is characterized by two parameters: the metal strips width $a$ and IDT period $b$, giving a duty ratio $\eta=a/b$. Because the size of LN waveguide is small, the two parameters have great influence on the resonant frequency of IDT as well as the conversion efficiency. Figures~\ref{fig:4}(b) and (c) show the electrical impedance of the IDT at different frequency when quasi-Love and quasi-Rayleigh mode is excited, respectively. Here, the waveguide is along $Y$ axis of LN, and IDT is made by $50\,\mathrm{nm}$-thick aluminum with $\eta=0.5$. The numerical results indicate that $b=730\,\mathrm{nm}$ and $787\,\mathrm{nm}$ matche the quasi-Love and quasi-Rayleigh mode at $6\,\mathrm{GHz}$, respectively. The electric impedance curves show asymmetric features that consists of  a peak and a dip~\cite{Lakin1992_IDT}, which represent the parallel resonance ($f_{p}$) and series resonance ($f_{s}$) of the structure according to the Butterworth-Van Dyke model~\cite{Schmid1991_IDT}. The effective electromechanical coupling coefficient ($k_{\mathrm{eff}}^{2}$) can be directly estimated from the two frequencies through $k_{\mathrm{eff}}^{2}\approx\frac{\pi^{2}}{4}(1-f_{s}/f_{p})$. Our calculations give $k_{\mathrm{eff}}^{2}=0.133$ for quasi-Love mode but only $5.76\times10^{-6}$ for the quasi-Rayleigh mode. The reason behind the more efficient coupling is that the coupling between the quasi-Love mode and the IDT electric field in our structure utilizes the strongest piezoelectric tensor component. 

\section{Qubit-phonon Strong Coupling}

In addition to device design, the realization of strong coupling in our configuration is vital for further development of the cavity QAD system. In the following, we analyze the coupling strength between the transmon and the acoustic microring resonator. The energy of a piezoelectric phononic cavity mode is composed by 3 parts, including kinetic energy, strain energy and the electrostatic energy with the quasistatic approximation of the electric field is applied in the description of the piezoelectric effects~\cite{royer1999elastic_U}.
\begin{equation}
U=\frac{1}{2}\intop_{V}\rho vv^{*}dV+\frac{1}{2}\intop_{V}S_{I}T_{I}^{*}dV+\frac{1}{2}\intop_{V}E^{T}\cdot, D^{*}dV\label{eq:1}
\end{equation}
where $\rho$ is the density of materials, $v$ is the velocity fields, $S$ and $T$ are the strain and stress tensors, and $E$ and $D$ are the electric field and displacement field, respectively. The piezoelectric interaction can be described by the piezoelectric stress equations with
\begin{equation}
\begin{array}{c}
T=c\colon S-e^{T}\cdot E,\\
D=\epsilon\cdot E+e:S,
\end{array}\label{eq:2}
\end{equation}
where $e$ is the piezoelectric-stress coefficients, $c$ is the elastic-stiffness coefficients, $\epsilon$ is the permittivity of materials.

By introducing the quantization of phononic mode with bosonic operators $a^{\dagger}$, the strain and the electrostatic field operator of the phononic mode can be represented by~\cite{scully1997quantum_qo}
\begin{equation}
\begin{array}{c}
\boldsymbol{S}(r,t)=i\xi a^{\dagger}S^{\dagger}(x,y)e^{-i(m\varphi-\omega t)}+h.c.\\
\boldsymbol{E}_{i}(r,t)=i\xi aE_{i}(x,y)e^{i(m\varphi-\omega t)}+h.c.,
\end{array}\label{eq:3}
\end{equation}
where $\xi=(\frac{\hbar\omega}{\intop_{V}[\rho v_{i}(x,y)v_{i}^{\ast}(x,y)+S_{i}(x,y)T_{i}^{*}(x,y)+E_{i}^{T}(x,y)\cdot D_{i}^{*}(x,y)]dV})^{1/2}$ is the normalizaiton factor. Similarly, the microwave mode of transmon qubit can also be quantized with bosonic operators $b^{\dagger}$, then the interaction energy between transmon and phononic mode $U_{I}=\frac{1}{2}\intop_{V}dV\cdot S^{\dagger}\cdot e^{T}\cdot E$, where $E$ is the electric field generated by the coupling capacitance, can be represented with
\begin{align}
\hat{H}_{I} & =i\hbar g(a+a^{\dagger})(b-b^{\dagger}),\label{eq:4}
\end{align}
where $g$ is the single phonon coupling strength with
\begin{align}
g & =\frac{1}{2}\intop_{V}dV\cdot S_{zpf}^{\dagger}\cdot e^{T}\cdot\vec{\zeta}(r)\frac{4E_{c}}{e}n_{\mathrm{zpf}},\nonumber \\
 & \approx-\intop_{V}dV\cdot\frac{\xi}{2\hbar}S^{\dagger}(r)\cdot e^{T}\cdot\vec{\zeta}(r)\frac{4E_{c}}{e}\cdot\sqrt{\frac{1}{2}}(\frac{E_{J}}{8E_{c}})^{1/4},\label{eq:5}
\end{align}
where $n_{zpf}$ is the charge zero-point fluctuations, $\vec{\zeta}(r)$ is the electric field distribution around the coupling capacitor when $1\,\mathrm{V}$ biased voltage is applied to the coupling capacitor, $E_{J}$ and $E_{c}$ are the Josephson energy and the charging energy in transmon, respectively. In Eq.~(\ref{eq:5}), $\xi$, $\vec{\zeta}(r)$ and $S^{\dagger}(r)$ can be calculated numerically by the finite-element method, as well as the capacitance of the coupling IDT. $E_{J}$ and $E_{c}$ are determined by the working frequency of transmon (here we set as 6\,GHz)~\cite{Devoret2004_ScQBit}, then the coupling strength can be calculated. With the condition of $E_{J}\gg E_{c}$, the total Hamiltonian of the coupling system is~\cite{Xiang2013_Hybrid}
\begin{align}
H & =(\sqrt{8E_{c}E_{J}}-E_{c})a^{\dagger}a-\frac{E_{c}}{2}a^{\dagger}a^{\dagger}aa\nonumber \\
 & +\hbar\Omega b^{\dagger}b+i\hbar g(b+b^{\dagger})(a-a^{\dagger})\label{eq:6}
\end{align}

\subsection{Single phonon Coupling Strength}

\begin{figure}[t]
\centering{}\includegraphics[width=1\columnwidth]{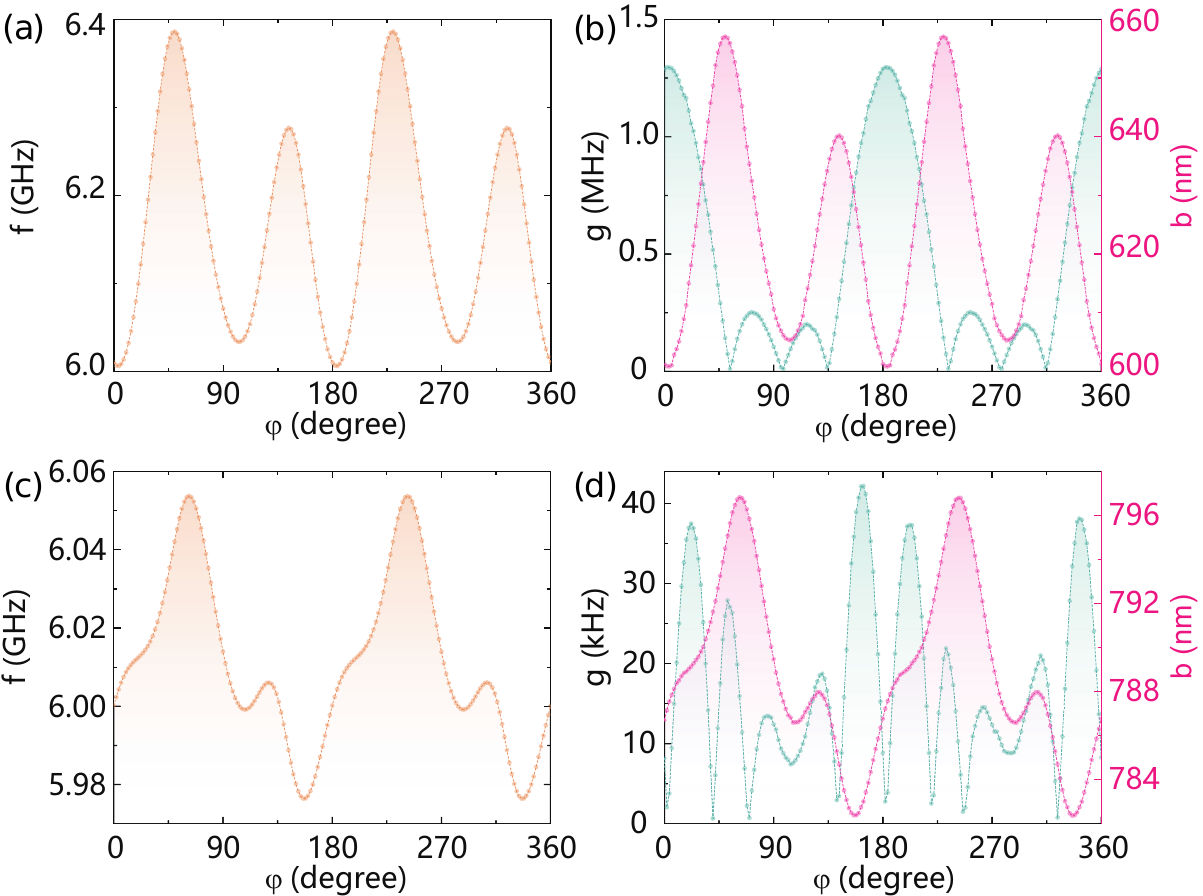}\caption{\label{fig:5}
(a) and (c) are the resonant frequency of IDT versus the location of it with quasi-Love and quasi-Rayleigh mode coupling, respectively. The red lines in (b) and (d) are the adjusted IDT period to ensure that the resonance frequency of the IDT is always 6 GHz with quasi-Love and quasi-Rayleigh mode coupling, respectively.The green lines are the corresponding single phonon coupling strength versus the location of the IDT. }
\end{figure}

As discussed in Fig.~\ref{fig:3}, the crystal orientation of the waveguide changes with the azimuthal angle $\varphi$ of a microring, and the coupling between IDT and microring varies at different $\varphi$. Figure~\ref{fig:5}(a) shows the resonant frequency of the coupling IDT for coupling the quasi-Love cavity mode. Starting from resonant frequency at $6\,\mathrm{GHz}$ when $\varphi=0$, as is designed in Fig.~\ref{fig:4}(c), the frequency sweeps up to $6.4\,\mathrm{GHz}$ as the waveguide orientation rotates around the microring, with $\varphi$ varying from 0 to 360 degrees. The period of the resonant frequency is also 180 degree similar to that of propagation constant shown in Fig.~\ref{fig:3}(a). To reveal the relationships between the single phonon coupling strength and the $\varphi$, the resonant frequency of IDT placed at different azimuthal angle is fixed to $6\,\mathrm{GHz}$ by adjusted the period of the IDT. The red curve in Fig.~\ref{fig:5}(b) shows the adjusted period of IDT at different $\varphi$. The maximum modification needs to approach $60\,\mathrm{nm}$, which highlights the careful device design.

The green curve in Fig.~\ref{fig:5}(b) shows the single phonon coupling strength according to Eq.~(\ref{eq:5}), with the radius of the microring is set as $50\,\mu m$. With only one pair of IDT fingers, we found that the largest coupling strength $g/2\pi=1.3\,\mathrm{MHz}$ is achieved when IDT align along Y-axis of LN. We have the linear scaling $g\propto N_{IDT}$ when the coupling IDT contributed capacitance to the transmon is negligible, given the number of IDT finger pairs $N_{IDT}$ is small. With just 10 finger pairs, it is anticipated that $g/2\pi>10\,\mathrm{MHz}$ is achievable, which certainly exceeds typical decay rates of transmon and the phononic modes~\cite{Xu2022}. This enters the strong coupling regime where coherent quantum state exchange dominates over decoherence. Remarkably, at special crystal orientations with $\varphi=54^\circ$, $98^\circ$, and $135^\circ$, the coupling strength vanishes completely. These off-coupling conditions occurs when the strain field become orthogonal to the relevant piezoelectric tensor components. These conditions might find applications for specific device designs that isolate certain modes from the microring.

Similarly, as is shown in Fig.~\ref{fig:5}(c), the resonant frequencies of the coupling IDT varies with $\varphi$ for the quasi-Rayleigh mode. Since the maximum frequency shift is only about 50\,kHz, leading to a much smaller modification ($8\,\mathrm{nm}$) of the IDT periods when fixing the IDT resonant frequency at 6\,GHz, as shown by the red line in Fig.~\ref{fig:5}(d). With maximum coupling of only $42\, \mathrm{kHz}$ at $\varphi=162^{\circ}$ and a mere $8\, \mathrm{kHz}$ along the Y-axis, which is 160 times weaker than that for quasi-Love. These results implies that for X-cut LN, the coupling between transmon and quasi-Rayleigh cavity modes is too weak to realize the strong coupling.

\subsection{Fabrication Tolerances}

\begin{figure}[t]
\centering{}\includegraphics[width=1\columnwidth]{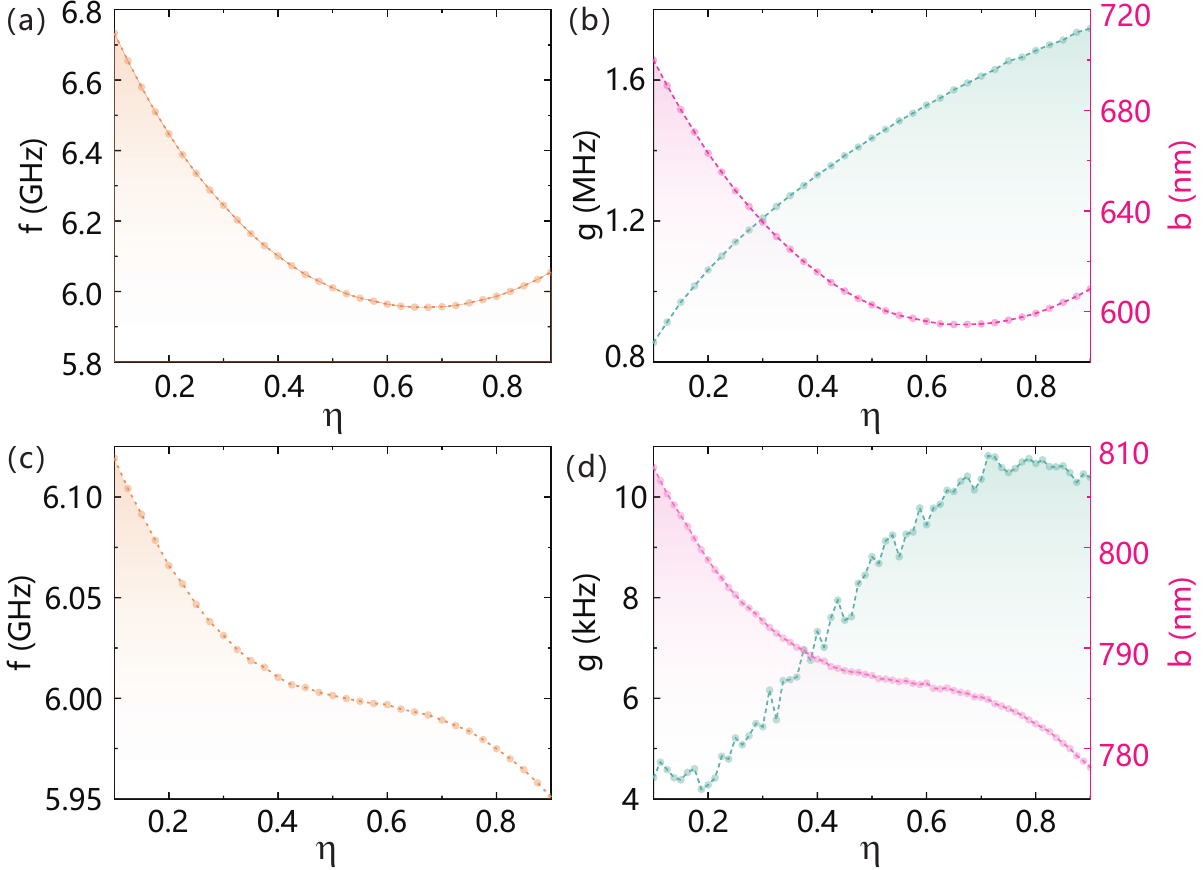}\caption{\label{fig:6}(a) and (c) are the resonant frequency of IDT versus the duty ratio of it with quasi-Love and quasi-Rayleigh mode coupling, respectively. The red lines in (b) and (d) are the adjusted IDT period to ensure that the resonance frequency of the IDT is always $6\,\mathrm{GHz}$ with quasi-Love and quasi-Rayleigh mode coupling, respectively. The green lines are the corresponding single phonon coupling strength versus the duty ratio of the IDT. }
\end{figure}

For practical fabrication processing, there are many fabrication imperfections need to be taken into account. The fabrication of coupling IDT is critical for realizing strong coupling. In particular, for the such a high  working frequency, the required IDT period of only $600\, \mathrm{nm}$, as shown in Fig.~\ref{fig:5}(c). The corresponding width of the finger is only about $150\, \mathrm{nm}$ when duty ratio is 0.5, imposing great challenges. Among the parameters of IDT, the thickness and the period of the finger pairs are relatively well controlled while the width of the fingers may deviate from the design value. Here we analyze how the inaccuracy of the duty ratio impact the performance of IDT.

Figure~\ref{fig:6}(a) displays the relationship between the resonant frequency of IDT and $\eta$ with quasi-Love mode coupling. We found that the resonant frequencies of the IDT is not only determined by the period but also by the duty ratio as well. This is because that the aluminum fingers induce three effects to change the resonance frequency, including mass loading, electrical shorting, and reflection of acoustic waves, with all effects are dependent on the $\eta$. When $\eta$ changes from 0.5 to 0.4, i.e. the width of the fingers increases by only $30\, \mathrm{nm}$, the resonant frequency will shift by $100\,\mathrm{MHz}$. The amount of frequency detuning is considerably large, as it is comparable with 3\,dB bandwidth of the IDT that consists of only $N_{p}=20$ pairs of finger, i.e., $\mathrm{BW_{3dB}}\approx0.885f_{0}/N_{p}\approx265\,\mathrm{MHz}$, where $\mathrm{N_{p}}$ is the number of period of IDT. In comparison, the coupling strength is relatively insensitive to the changing of $\eta$. As shown in Fig.~\ref{fig:6}(b), the $g$ only varies by 7\% when $\eta$ decrease by 0.1.

For quasi-Rayleigh mode, due to the weak interaction, the shift of resonant frequency of IDT only about 3\% even if the $\eta$ changes from 0.1 to 0.9 (Fig.~\ref{fig:6} (c)). However, Fig.~\ref{fig:6} (d) shows that the coupling strength has increased more than two-fold from $4\,\mathrm{kHz}$ to $11\, \mathrm{kHz}$, even though the absolute values are negligible. So for quasi-Love mode coupling, the duty ratio is a key parameter that needs to be strictly controlled during the design and fabrication process.

\section{Discussion}

The proposed hybrid cavity QAD platform opens up a wide range of opportunities for quantum information processing and the study of novel quantum phenomena. Here, we discuss three key aspects that our platform can be extended and applied.

First, the realization of quantum phononic circuits. The PnICs in the LNoS platform enable the coherent manipulation and routing of traveling phononic wave packets, which can serve as quantum information carriers. By treating transmon qubits as deterministic single-phonon sources and detectors, quantum phononic circuits can be realized. For example, Hong-Ou-Mandel (HOM) interference~\cite{Qiao2023_Hom}, a key quantum effect demonstrating the indistinguishability of single phonons, can be observed in a phononic beamsplitter circuit. The demonstration of such quantum phenomena in the phononic domain paves the way for the development of advanced quantum information processing protocols and algorithms based on traveling phononic qubits.

Second, scalable superconducting quantum processors via phonon-mediated interactions. Phonons in the LNoS platform can mediate long-range interactions between superconducting qubits, enabling the scaling up of quantum processors. By coupling multiple transmon qubits to a common phononic mode, such as a microring resonator, entanglement between distant qubits can be generated and maintained~\cite{Chou2025_entanglement}. This approach circumvents the limitations of nearest-neighbor coupling in conventional superconducting qubit architectures. Furthermore, the multimode nature of the phononic resonators allows for the exploration of multimode quantum QAD, where multiple phononic modes are exploited for encoding and manipulating quantum information~\cite{Han2016_multimode,Moores2018_multimode}. The study of giant atom effects, where qubits couple to phononic modes at multiple points, can also be pursued in this platform, leading to novel quantum dynamics and enhanced qubit-phonon interactions~\cite{Andersson2019}.

Third, hybrid photonic-phononic-superconducting circuits~\cite{Xu2022_chip}. The LNoS-QAD platform can be further extended by integrating it with photonic circuits, leveraging the strong Brillouin interaction in the waveguides~\cite{Yang2024a_sbs,Neijts2024_sbs}. Brillouin scattering, which involves the coherent coupling between optical photons and acoustic phonons, enables microwave-to-optical quantum frequency conversion~\cite{Yang2024_sbs}. This process allows the transfer of quantum states between superconducting qubits operating in the microwave domain and optical photons suitable for long-distance communication. By incorporating Brillouin-active waveguides into the LNoS platform, efficient and low-noise quantum frequency converters can be realized. This hybrid approach enables the development of quantum networks, where superconducting quantum processors are connected via optical links, facilitating the distribution of entanglement and the realization of large-scale quantum computing systems.

\section{CONCLUSION}
A scalable cavity QAD platform based on LNoS chip is proposed. This platform leverages the superior coherence properties of superconducting qubits fabricated on the single-crystal sapphire substrate, while low-loss phonons are efficiently confined and guided in the PnICs. The strong piezoelectric effect of LN allows for robust coupling between qubits and acoustic waves, enabling the strong coupling regime with just a few interdigital transducer electrodes on the phononic microring resonator. This versatile platform promises the realization of quantum phononic circuits, scalable superconducting quantum chip with phonon-mediated interactions, and hybrid quantum systems through Brillouin interactions with photonic circuits. We anticipate that this platform will unlock exciting new opportunities in solid-state quantum technologies.

\begin{acknowledgments}
This work was supported by the National Natural Science Foundation of China (Grants No.~92265210, 92165209,  92365301, 12104441, 123B2068, 12474498, and 11925404), Innovation Program for Quantum Science and Technology (Grant Nos. 2021ZD0300200 and 2024ZD0301500). This work was also supported by the Fundamental Research Funds for the Central Universities and USTC Research Funds of the Double First-Class Initiative. C.-L.Z. was also supported by Beijing National Laboratory for Condensed Matter Physics (2024BNLCMPKF007). The numerical calculations in this paper have been done on the supercomputing system in the Supercomputing Center of University of Science and Technology of China. This work was partially carried out at the USTC Center for Micro and Nanoscale Research and Fabrication.

\end{acknowledgments}

\end{document}